\documentclass[aps,pra,superscriptaddress,twocolumn,a4paper,notitlepage]{revtex4-1}

\usepackage{amssymb}
\usepackage{amsmath}
 \usepackage{color}
\usepackage{psfrag}
\usepackage{ifpdf}
\ifpdf
\usepackage{epstopdf}   
\fi
\usepackage[raggedright]{titlesec}

\newcommand{\ket}[1]{\ensuremath{\vert#1\rangle}}
\newcommand{\bra}[1]{\ensuremath{\langle #1\vert}}
\newcommand{\bk}[2]{\ensuremath{\langle #1\vert #2\rangle}}
\newcommand{\kb}[2]{\ensuremath{\vert #1 \rangle \langle #2 \vert}}
\newtheorem{theorem}{Theorem}
\newtheorem{defin}{Definition}

\newcommand{\tr}{\ensuremath{\mathrm{tr}}}

\def\id{{\mathchoice {\rm 1\mskip-4mu l} {\rm 1\mskip-4mu l} {\rm 1\mskip-4.5mu l} {\rm 1\mskip-5mu l}}}

\definecolor{SecCol}{RGB}{120, 120, 120}
 \titleformat{\section}
 {\bfseries  }{\thesection}{.5em}{}
 \renewcommand\thesection{\arabic{section}}

\begin{document}

\title{Decoherence in open Majorana systems}

\author{Earl T.\ Campbell}
 \affiliation{Department of Physics \& Astronomy, University of Sheffield, Sheffield, S3 7RH, United Kingdom.}
 \email{earltcampbell@gmail.com}

\begin{abstract}

Coupling to a thermal bath leads to decoherence of stored quantum information.   For a system of Gaussian fermions, the fermionic analog of linear or Gaussian optics, these dynamics can be elegantly and efficiently described by evolution of the system's covariance matrix.  Taking both system and bath to be Gaussian fermionic, we observe that decoherence occurs at a rate that is independent of the bath temperature.  Furthermore, we also consider a weak coupling regime where the dynamics are Markovian.  We present a microscopic derivation of Markovian master equations entirely in the language of covariance matrices, where temperature independence remains manifest. This is radically different from behaviour seen in other scenarios, such as when fermions interact with a bosonic bath.  Our analysis applies to many Majorana fermion systems that have been heralded as very robust, topologically protected, qubits.  In these systems, it has been claimed that thermal decoherence can be exponentially suppressed by reducing temperature, but we find Gaussian decoherence cannot be cooled away.
\end{abstract}

\maketitle

Thermalization through interaction with an external bath is one of the principal mechanisms by which quantum systems lose information.  In quantum technologies, rapid thermalisation destroys their advantage over classical counterparts. By better understanding these processes, one hopes to identify and engineer physical systems that act as more robust stores of quantum information.  In topologically ordered systems, information is stored non-locally within the degenerate ground space of some large many-body system.  The primary benefit of topology is robustness against random adiabatic fluctuations in the system Hamiltonian.  Damage from such noise is exponentially suppressed with system size.  Topological systems also have an energy gap $\Delta$ between the degenerate ground space and excited states, and are said to be protected by the gap against thermal excitations.   A common claim~\cite{Nayak08} is that thermal processes occur at a rate $e^{-\Delta / T}$, which is sometimes called the Arrhenius law.  The bold conclusion is that topology can exponentially eliminate noise merely by increasing system size and decreasing temperature.

Of all topological systems, Majorana zero modes have attracted the most attention. It was theorized that a so-called Kitaev wire supports Majorana zero modes at edges, which could be realised in simple solid state hetrostructures~\cite{Alicea12}, for example a nanowire coupled to a conventional s-wave superconductor~\cite{Kitaev01}.   This drove a series of experiments, eventually leading to observations of Majorana edge modes~\cite{Mourik12,Nadj14,Franz13}.  Beyond topological robustness, Majorana zero-modes also possess the braiding statistics of non-Abelian Ising anyons.  Though insufficient for direct quantum computation, braiding Ising anyons can demonstrate nonlocality, teleportation and superdense coding~\cite{Campbell14}.  Furthermore, Ising anyon braiding can be promoted to full quantum computing when supplemented with some nontopological (noisy) operations~\cite{Bravyi06,deMelo13}.  

\begin{figure*}[t]
    \includegraphics{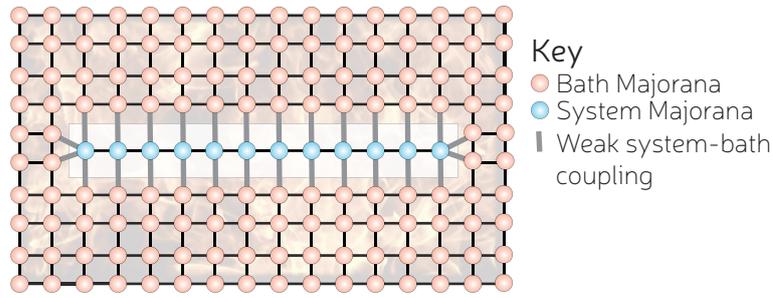}
    \psfrag{X}{Bath Majorana}
    \caption{A chain of cool Majorana fermions weakly coupled to a large thermal 2-dimensional bath of Majorana fermions.  This provides an example of the general paradigm we work within, though our results apply to all Gaussian systems.}
    \label{FigWireBath}
\end{figure*}

The physics of these Majorana systems is especially tractable as their Hamiltonians are quadratic in fermion creation and annihilation operators.  We say such a system is Gaussian, or quasifree fermionic, in analogy with Gaussian linear optics.  Gaussian states can be described purely in terms of the expectation value of quadratic observables, which are captured by a covariance matrix.  Furthermore, some dissipative processes can be described within this powerful covariance matrix formalism~\cite{Bravyi05b,Prosen08,Prosen10,Prosen10b,eisert10,Bravyi11,bernigau13}, and allow single fermions to hop between system and bath via $a^{\dagger}_{S}a_B$. Single fermion hopping  violates conservation of fermion parity in the system, which is otherwise respected by unitary evolution. It is a toxic process that can cause errors without creating excitations, circumventing arguments that energy penalties suppress thermal processes to a rate $e^{-\Delta / T}$.  In particular, Majorana modes in the Kitaev wire (see Fig.~\ref{FigWireBath}) have been shown to decohere due to fermion hopping at rates independently of system size or the system gap~\cite{Budich12,Mazza13}, and we will review these results in detail.  This article considers all Gaussian fermionic systems, not just the Kitave wire, and how they decohere as a function of temperature.  A single fermion appearing in the system will have a partner appear in the environment, and so perhaps there is hope that a gapped bath Hamiltonian will provide an energy penalty inhibiting these processes. 

We present a very general, yet simple, argument that decoherence is independent of temperature, assuming only that the system-bath is governed by a Gaussian Hamiltonian.  We extend this argument by providing a microscopic derivation of a master equation in the weak coupling regime, and again observe temperature independent decoherence. When fermions couple to bosonic baths, or through quartic fermion-fermion interactions $a^{\dagger}_{S}a_S a^{\dagger}_B a_B$, one would instead find a non-trivial temperature dependence. Any real physical system will experience noise from multiple sources, mostly with temperature dependent rates.  However, fermionic hopping presents a constant background noise that cannot be suppressed through cooling.  This adds to a growing body of work~\cite{Goldstein11,Konschelle13,Rainis12,Pedrocchi15} that shows the outlook for Majorana fermions makes them less promising quantum memories than initially supposed.  Our conclusions can be avoided by going beyond Gaussian fermions, for instance by making use of complex many-body interactions used in the passive quantum memories reviewed in Ref.~\cite{Brown14}.  We discuss how our results demonstrate a break down of the Arrhenius law, whilst still satisfying a notion of detailed balance.  Decoherence of two-level systems, such as spins, has been studied when they couple to spin or fermion baths~\cite{ku05,Shnirman05} where a temperature dependence is observed but at low temperatures relaxation rates plateau, similarly breaking the Arrhenius law.  Throughout, we use the phrase thermalize as synonymous with equilibrate or approach steady state. The reader should not infer that the steady state is the thermal Gibbs distribution with respect the system Hamiltonian and ambient temperature, as may not be the case.


\section{Covariance Matrix Formalism}
Here we present standard techniques for working with Gaussian fermions~\cite{Bravyi05b,Prosen08,Prosen10,Prosen10b,eisert10,Bravyi11,bernigau13}, and use them to show that decoherence is independent of bath temperature.  Relaxation of Gaussian fermionic open systems has be studied in detail (see e.g. Refs.~\cite{kastoryano13,Temme14}, but these did not include a model of the bath as also composed of Gaussian fermions.  The first step is to map $n$ Dirac fermions (e.g. electrons) with annihilation and creation operators $\{ a_n, a_n^\dagger \}$ into $2n$ Majorana operators
\begin{equation}
    c_{2n-1}=a_n +  a_n^{\dagger} \mbox{ ,    } c_{2n}=i(a_n - a_n^{\dagger}) .
\end{equation}
They are still fermionic in anti-commutation $\{ c_{j}, c_{k} \} = \delta_{j,k}$, but differ from Dirac fermions in that they satisfy $c_{j}^{\dagger}=c_{j}$ and $c^2_{j}=\id$.  For any quantum state $\rho$, the covariance matrix has real elements composed of second moments
\begin{equation}
    \Gamma_{j,k} =\frac{i}{2} \tr[ ( c_{j} c_{k} - c_{k}c_{j}) \rho ] .
\end{equation}
Due to fermion anticommutation, the covariance matrix is skew-symmetric $\Gamma^{T}=-\Gamma$.  Because of conservation of fermion parity, first moments always vanish $\tr( c_{j} \rho )=0$.  For Gaussian states, expectation values of higher moments are determined by the covariance matrix via Wick's theorem.  Likewise a quadratic Hamiltonian $\hat{H}$ is described by a matrix $H$ so that
\begin{equation}
\label{EqQuadHam}
    \hat{H} = \frac{i}{4}\sum_{j,k}  H_{j,k} c_{j} c_{k} .
\end{equation}
Again, $H$ must be real for the Hamiltonian to be Hermitian, and furthermore $H$ can always be chosen skew-symmetric $H=-H^T$.  For a closed quantum system evolving unitarily, the covariance matrix evolves according to
\begin{equation}
    \frac{d \Gamma(t) }{d t} =[  \Gamma(t) , H],
\end{equation}
where $[ \cdot , \cdot ]$ is the commutator. For a time independent Hamiltonian, this results in 
\begin{equation}
    \Gamma(t) = e^{H t} \Gamma(0) e^{H^T t}.
\end{equation}
A joint system-bath covariance matrix has the form
\begin{equation}
    \Gamma = \left( \begin{array}{cc} \Gamma_{S} & -\Gamma_C^{T} \\    \Gamma_C & \Gamma_{B} \end{array} 
\right) ,
\end{equation}
where $\Gamma_{S}$ and $\Gamma_{B}$ represent, respectively, the system and bath covariance matrices, and $\Gamma_C$ records system-bath correlations.  In other words, given a state $\rho$ with covariance matrix $\Gamma$, tracing out the bath gives a reduced density matrix $\tr_{B}(\rho)$ with covariance matrix $\Gamma_S$.  We define $\{ \cdots \}_B$ to denote this process of reducing the covariance matrix, so $\{ \Gamma \}_B = \Gamma_S$.   In general, the reduced covariance matrix of an open quantum system will be
\begin{equation}
    \Gamma_{S}(t) = \{ e^{Ht} \Gamma(0) e^{H^{T}t} \}_B .
\end{equation}
For an uncorrelated system $\Gamma_C=0$, the covariance matrix has a direct sum form $ \Gamma = \Gamma_S \oplus \Gamma_B$.  The direct sum is linear, so that uncorrelated states have the form 
$\Gamma = (\Gamma_S \oplus \mathbf{0}) + (\mathbf{0} \oplus \Gamma_B)$, where $\mathbf{0}$ denotes an all zero matrices.  The covariance reduction $\{ \cdots \}_B$ is also linear, and so uncorrelated states evolve to 
\begin{equation}
    \Gamma_{S}(t) = \{ e^{Ht} (\Gamma_S \oplus \mathbf{0}) e^{H^{T}t} \}_B + \{ e^{Ht} (\mathbf{0} \oplus \Gamma_B) e^{H^{T}t} \}_B .
\end{equation}
Notice that the first term is independent of the bath variables such as temperature, and can be more compactly written as
\begin{equation}
    \{ e^{Ht} (\Gamma_S \oplus \mathbf{0}) e^{H^{T}t} \}_B  = D(t)\Gamma_S D(t)^T,
\end{equation}
where $D(t) := \{ e^{Ht} \}_B$.  We are interested in the rate of decoherence. How quickly will two states become indistinguishable?  Consider two different initial covariance matrices $\Gamma^{\tilde{\rho}}=\Gamma^{\tilde{\rho}}_S \oplus \Gamma_B$ and  $\Gamma^{\rho}=\Gamma^{\rho}_S \oplus \Gamma_B$, e.g. describing logical encodings of qubit states.  The time evolved difference between these covariance matrices is
\begin{equation}
    \delta(t)  :=  \Gamma_{S}^{\rho}(t)-  \Gamma_{S}^{\tilde{\rho}}(t)  =   D(t) ( \Gamma_{S}^{\rho}-  \Gamma_{S}^{\tilde{\rho}} ) D(t)^T .
\end{equation}
We observe that this is entirely independent of the bath temperature.  As $\delta(t) \rightarrow \mathbf{0}$, the states becomes indistinguishable.  Using $|| \cdots ||$ to denote the operator norm (the largest singular value) of a matrix, it is straightforward to show
\begin{equation}
    \tr[ i \tilde{c}_{j}\tilde{c}_{k} (\rho -\tilde{\rho})] \leq || \delta(t) || ,
\end{equation}
where $ \tilde{c}_{j}$ and $\tilde{c}_{k}$ are any pair of anti-commuting Majorana operators. Therefore, small $|| \delta(t) ||$ entails low probability of distinguishing $\rho$ and $\tilde{\rho}$ through a single Gaussian measurement.  We show later that this statement can be extended to completely general measurements.  The operator norm is submultiplicative and transpose invariant so that 
\begin{equation}
     || \delta(t) ||  \leq || D(t) ||^2 ||( \Gamma_{S}^{\rho}-  \Gamma_{S}^{\tilde{\rho}} )  || \leq 2  || D(t) ||^2,
\end{equation}
with smaller $|| D(t) ||$ entailing more decoherence.  We have used $||( \Gamma_{S}^{\rho}-  \Gamma_{S}^{\tilde{\rho}} )  ||\leq 2$ to present an upperbound that is also independent of the initial state.  Without system-bath interactions $D(t)=  e^{H_S t}$ is unitary so that  $|| D(t) ||=1$, but interactions lead to dissipation and $|| D(t) ||<1$.  Under very general conditions we have determined that Gaussian decoherence occurs, quite remarkably, independently of the bath temperature. In some instances $|| D(t) ||$ may decrease with time, only to revive later.   However, for a sufficiently complex bath we expect Markovian behavior lead to exponentially fast decoherence $|| D(t) ||= e^{-\lambda t} $ for some $\lambda$.   Next we introduce the formalism of Gaussian fermionic master equations, and then proceed to perform a microscopic derivation assuming weak-coupling.  In such derivations various approximations are made, yet we find they respect temperature invariance.
 
\section{Gaussian fermionic master equations}
Here we review Gaussian fermionic master equations following Refs.~\cite{Bravyi05b,Prosen08,Bravyi11}.   The dynamical equation for such a system is
\begin{equation}
\label{EqGenQuantumChannel}
    \frac{d \Gamma(t) }{d t} = X \Gamma(t) + \Gamma(t) X^{T} + Y,
\end{equation} 
where, for dissipative dynamics, $X$ is not necessarily skew-symmetric.  In general, $X=-H-P$ where $H$ represents the unitary component (and is skew symmetric) and $P$ represents the dissipative part (and is symmetric and positive).  Typically, dissipative systems will have one steady state $\Gamma_{ss}$ that satisfies $X \Gamma_{ss} + \Gamma_{ss} X^{T} =- Y$, and then
\begin{equation}
    \Gamma(t) = e^{X t}(\Gamma(0) - \Gamma_{ss})e^{X^T t}  + \Gamma_{ss}.
\end{equation}
These very general dynamical equations are the covariance matrix representation of a class of Lindblad master equations of the form
\begin{equation}
    \frac{d \rho(t)}{dt} = i[H , \rho(t) ] + \sum_{\mu} (2 L_{\mu} \rho(t) L_{\mu}^{\dagger} - \{ L_{\mu}^{\dagger} L_{\mu}, \rho(t) \} ) ,
\end{equation}
where the Hamiltonian $H$ must be of the quadratic form introduced in Eq.~(\ref{EqQuadHam}) and the Lindblad operators are linear in Majorana operators
\begin{equation}
    L_{\mu} = \sum_{j} l_{\mu,j} c_{j}.
\end{equation}
One can prove~\cite{eisert10,Bravyi11}  that such a master equation gives rise to a Gaussian quantum channel with 
\begin{eqnarray}
    X &=& -H - (M + M^*) ,\\
    Y &=& 4i(M + M^*) ,
\end{eqnarray}
where $M$ has elements $M_{j,k}=\sum_{\mu} l_{\mu,j}l^{*}_{\mu,k}$.  The matrices $X$ and $Y$ will always be real so that $X^T=X^\dagger$ and $Y^T=Y^\dagger$. In the literature, there is some discussion of how such systems decohere Majorana qubits.  However, prior work has centered on only the Kitave wire and has not considered how variables in Lindblad equations depend on the underlying system parameters. 

\section{Rates of thermalisation}
\label{sec_Convergence}

We quantify the decoherence by the operator norm of $\delta(t)  :=  \Gamma_{S}^{\rho}(t)-  \Gamma_{S}^{\tilde{\rho}}(t)$.  In general, $||\delta(t)||\leq 2 ||D(t)||^2$ and from the previous section we see that $D(t)=e^{Xt}$.  Theorem IX.3.1 of Bhatia~\cite{bhatia97} shows that for any $X$, we have $|| e^{Xt} || \leq || e^{(X+X^{\dagger})t/2} ||$ as previously employed in this setting by Bravyi-K\"{o}nig~\cite{Bravyi11}.   Using the shorthand $P:=(X+X^\dagger)/2$, this entails 
\begin{equation}
    || \delta(t)|| \leq 2|| e^{-Pt} ||^2 = 2e^{-2\lambda_{P}t} .
\end{equation}
We know $\lambda_{P}$ is non-negative, so provided $P$ does not have any zero eigenvalues there will be exponentially rapid decoherence.  It is well known that decoherence rates are governed by the spectral properties of $P$.  Our contribution is investigation of the dependence of this spectrum on microscopic factors.

The above statement holds for any pair of Gaussian states and any Gaussian channel.  We are also interested in the more specific scenario where the Gaussian states lie in the degenerate groundspace of a physical Hamiltonian.  Eigenstates of this Hamiltonian can be simultaneously diagonalised, and similarly their covariance matrices can be simultaneously brought into Williamson normal form.  There exists a orthogonal matrix $O$, such that for every pure Gaussian $\psi$ that is a eigenstate of the Hamiltonian we have
\begin{equation}
    \Gamma^{\psi} = O \left( \bigoplus_j \gamma^\psi_j \tilde{\rho} \right) O^{T} ,
\end{equation}
where  $\gamma^\psi_j \in \{ \pm 1 \}$ distinguish different eigenstates, and
\begin{equation}
    \tilde{\rho} = \left( \begin{array}{cc} 0 & 1 \\
    -1 & 0  \end{array} \right).
\end{equation}
For some values of $j$, the numbers $\gamma^\psi_j$ tell us whether there is an excitation present.  However, for degenerate Hamiltonians there are a set of degeneracy indices $G$, such that $\gamma^\psi_j$ can vary without creating excitations for all $j \in G$.   Therefore, the covariance matrices of groundstates break up into two blocks $\Gamma^{\psi}= O[\Gamma^{\psi}_{G} \oplus \Gamma_{E}]O^T$ where $\Gamma_{E}$ is the same for all groundstates, and 
\begin{equation}
    \Gamma^{\psi}_{G} = \bigoplus_{j\in G} \gamma^\psi_j \tilde{\rho} .
\end{equation}
Let us consider two encoded ground states $\ket{\psi}$ and $\ket{\phi}$.  We deduce that $\delta=\Gamma^{\psi}-\Gamma^{\phi}=O [(\Gamma^{\psi}_G- \Gamma^{\phi}_G)  \oplus \mathbf{0}] O^T$ where $\mathbf{0}$ is a zero matrix.  

The matrix representing Hamiltonian dynamics has the same block structure as the covariance matrices, so that $H = O [H_G \oplus H_E] O^T$.
When such a system is exposed to an environment, its dynamics are dictated by some matrix $X$.  In many situations (including weakly coupled Markovian systems), $X$ will obtain the same block structure as $H$, so that $X= O [ H_G \oplus X_E ] O^T$.  It follows that 
\begin{equation}
    \delta(t) := \Gamma^{\psi}(t)-\Gamma^{\phi}(t) = O[ e^{X_{G}t} (\Gamma^{\psi}_G- \Gamma^{\phi}_G) e^{X_{G}^T t} \oplus \mathbf{0} ]O^T .
\end{equation}
Defining $P_G=(X_G+X_G^{\dagger})/2$, and using the same arguments as earlier we find
\begin{equation}
\label{deltaCon}
    || \delta(t) || \leq e^{-2 \lambda_{P_G}t} || (\Gamma^{\psi}_G- \Gamma^{\phi}_G) || \leq 2e^{-2 \lambda_{P_G}t}.
\end{equation}
Therefore, the decoherence of encoded ground states is governed by the spectrum of $P_G=-(X_G+X_G^{\dagger})/2$.  

The above arguments tell us that two initial Gaussian states undergoing Markovian dynamics will exponentially converge towards having identical covariance matrices.  Therefore, the probability of distinguishing these states through a single Gaussian measurement decreases exponentially in time.  However, a non-Gaussian measurement or multiple Gaussian measurements could prove more successful. In general, any strategy for distinguishing two states can always be captured by an observable $M$ with eigenvalues $\pm 1$, with an average success probability
\begin{equation} 
\mathrm{Pr}=\frac{1}{2}(1+\mathrm{tr}[ M (\rho - \tilde{\rho}) ]).
\end{equation}
It is well known that $\mathrm{Pr} \leq  \frac{1}{2} ||\rho - \tilde{\rho} ||_{\mathrm{tr}}$ where the trace norm is $|| A ||_{\mathrm{tr}}  := \mathrm{tr} [ \sqrt{A^{\dagger} A} ]$.  Therefore, we aim to show convergence in 1-norm. We again compare two initial pure states encoding a qubit into 4 Majorana modes, and find that the time evolved density matrices states $\rho(t)$ and $\tilde{\rho}(t)$ satisfies
\begin{equation}
\label{TRnormCon}
    ||\rho - \tilde{\rho} ||_{\mathrm{tr}} \leq 2 e^{-2 \lambda_{P_G}t},
\end{equation}
This follows quickly from Eq.~(\ref{deltaCon}) as we show in App.~\ref{AppTRnorm}.

\section{Derivation of master equation}
In this section we present a weak-coupling derivation of a Gaussian master equation in the covariance matrix formalism.  Many of the steps directly mirror those made in a textbook density matrix derivation (see e.g. Chap 3 of Ref.~\cite{petruccione02}). We assume both the system, the heat bath and their interaction is entirely Gaussian.  In addition, we make the usual assumptions involved in deducing master equations, notably that system-bath coupling is weak and that the system-bath are effectively uncorrelated at all times.  
The whole system-bath dynamics are described by a Hamiltonian with block matrix structure
\begin{equation}
    H = \left( \begin{array}{cc} H_{S} & -H^{T}_{I} \\    H_{I} & H_{B} \end{array} 
\right) ,
\end{equation}
where $H_{S}$ and $H_{B}$ represent, respectively, the system and bath Hamiltonians and satisfy $H_{x}=-H_{x}^{T}$ for $x=S,B$.  The interaction is represented by $H^{T}_{I}$ a real-valued, not necessarily square, matrix.  The initial ($t=0$) system-bath convariance matrix has the form
\begin{equation}
    \Gamma(0) = \left( \begin{array}{cc} \Gamma_{S}(0) & 0    \\
    0 & \Gamma_{B} \end{array} \\ \right) .
\end{equation}
Before proceeding we shift to an interaction picture, defining $\Gamma_{\mathrm{int}}(t):= e^{(H_{S} \oplus H_{b})t} \Gamma(t) e^{-(H_{S} \oplus H_{b})t}$.  It follows that
\begin{equation}
   \frac{d \Gamma_{\mathrm{int}}(t)}{dt} = [  \Gamma_{\mathrm{int}}(t) , H_{\mathrm{int}}(t) ] ,
\end{equation}
where 
\begin{equation}
 H_{\mathrm{int}}(t) = 
 e^{(H_S \oplus H_B)t}  \left( \begin{array}{cc} 0 & -H_I \\
 H_I & 0 \end{array} \right)  e^{-(H_S \oplus H_B)t} .
\end{equation}
This simplifies to
\begin{equation}
 H_{\mathrm{int}}(t) = \left( \begin{array}{cc} 0 & -H_{I}^{T}(t) \\
 H_{I}(t) & 0 \end{array} \right) ,
\end{equation}
where $H_{I}(t)= e^{H_{B}t} H_{I} e^{-H_{S}t}$.  Once in the interaction picture, we integrate over time to find
\begin{equation}
    \Gamma_{\mathrm{int}}(t) = \Gamma_{\mathrm{int}}(0) + \int_{0}^{t} [\Gamma_{\mathrm{int}}(s), H_{\mathrm{int}}(s)  ] ds .
\end{equation}
Therefore the time derivative is 
\begin{equation}
\begin{split}
   \frac{d \Gamma_{\mathrm{int}}(t)}{dt} &= [ \Gamma_{\mathrm{int}}(0), H_{\mathrm{int}}(t)] \\ \nonumber
   &+ \int_{0}^{t} [ [\Gamma_{\mathrm{int}}(s), H_{\mathrm{int}}(s)  ] , H_{\mathrm{int}}(t)] ds .
\end{split}
\end{equation}
We are only interested in the system covariance matrix, which is the covariance reduction $\{...\}_{B}$ of the above expression.  It is straightforward to verify $\left\{ [\Gamma_{\mathrm{int}}(0), H_{\mathrm{int}}(t) ] \right\}_{B}=0$, so
\begin{equation}
 \frac{d \{ \Gamma_{\mathrm{int}}(t) \}_{B}}{dt} = \int_{0}^{t} \{ [ [  \Gamma_{\mathrm{int}}(s),H_{\mathrm{int}}(s)],H_{\mathrm{int}}(t)] \}_{B} ds .
\end{equation}
Next, we assume the coupling is weak and that the system stays uncorrelated from the bath.  Formally, this entails that $\Gamma_{\mathrm{int}}(s) \rightarrow \{\Gamma_{\mathrm{int}}(t)\}_{B} \oplus \Gamma_{B}$, and also that $H_{\mathrm{int}}(s) \rightarrow H_{\mathrm{int}}(t-s)$ and the integral is extended to infinity.  Such assumptions directly mirror those made on the level of Hilbert spaces and result in the expression
\begin{equation}
\label{EqDiffInt}
 \frac{d \tilde{\Gamma}(t)}{dt} = \int_{0}^{\infty} \{ [ [\Gamma_{\mathrm{int}}(t), H_{\mathrm{int}}(t-s)  ],H_{\mathrm{int}}(t)] \}_{B} ds ,
\end{equation} 
where have made use of the shorthand $\tilde{\Gamma}(t):=\{ \Gamma_{\mathrm{int}}(t)\}_{B}$.  Next we may use our knowledge of the block structure of the covariance matrices to evaluate the commutator, and find
\begin{multline}
\nonumber
    \{ [ [\Gamma_{\mathrm{int}}(t) , H_{\mathrm{int}}(t-s)  ] , H_{\mathrm{int}}(t) ] \}_{B}  = \\ -H_{I}^{T}(t)H_{I}(t-s)\tilde{\Gamma}(t) - \tilde{\Gamma}(t)H_{I}^{T}(t-s) H_{I}(t) \\
      + H_{I}^{T}(t)\Gamma_{B} H_{I}(t-s) + H_{I}^{T}(t-s)\Gamma_{B} H_{I}(t) .
\end{multline}
Combing this with Eq.~(\ref{EqDiffInt}), and collecting terms to match Eq.~(\ref{EqGenQuantumChannel}) we have
 \begin{equation}
 \frac{d \tilde{\Gamma}(t) }{dt} = X \tilde{\Gamma} + \tilde{\Gamma} X^{T} + Y ,
\end{equation}
where
\begin{eqnarray}
    X &=& -\int_{0}^{\infty} H_{I}^{T}(t)H_{I}(t-s) ds , \\ \nonumber
    Y &=& \int_{0}^{\infty} H_{I}^{T}(t)\Gamma_{B} H_{I}(t-s) + H_{I}^{T}(t-s)\Gamma_{B} H_{I}(t) ds .
\end{eqnarray}
We have succeeded in deriving a form of a Gaussian quantum channel.  Though to make these equations meaningful we require that the integrals converge to finite values. For finite size matrices the integrands will be periodic functions and typically do not converge to a finite value.  Whereas, in the limit of infinite matrices the eigenvalue spectrum may become continuous and the integrand may vanish in the large $s$ limit.  Furthermore, to yield Markovian dynamics the resulting $X$ and $Y$ must be time independent.  Before proceeding we can already observe that all $\Gamma_B$ dependence has vanished from $X$.

Presently, the matrix $X$ still carries an overt time dependence, which can be removed by making the secular approximation (SA).  First we recall the explicit time dependence, $H_{I}(t)=e^{H_B t} H_{I} e^{-H_S t}$ so that 
\begin{eqnarray} \nonumber
    X & = & \int_{0}^{\infty} e^{H_S t} H^T_{I} e^{-H_B t} e^{H_B(t-s)} H_I e^{-H_S(t-s)}  ds , \\
    & = & \int_{0}^{\infty} e^{H_S t} H^T_{I} e^{H_B s} H_I e^{-H_S(t-s)}  ds .
\end{eqnarray}
We proceed by noting that $H_{S}$ is real and skew-Hermitian, so it has imaginary eigenvalues $ i\omega_{j}$, eigenvectors $\ket{j}$, and a diagonal form
\begin{equation}
    H_{S} = i \sum_{j} \omega_{j} \ket{j}\bra{j} .
\end{equation}
This entails 
\begin{eqnarray}
\label{Eq_X_RWAed} \nonumber
    X &=& - \sum_{j,k}  \int_s^{\infty}  e^{i(\omega_{j}-\omega_{k})t}  \ket{j}\bra{k} f_{j,k}(s)  e^{i 
\omega_{k}s}, 
\end{eqnarray} 
where
\begin{equation}
   f_{j,k}(s)= \bra{j}  H_I^T e^{-H_B s} H_I \ket{k} .
\end{equation}
The SA asserts that terms with rapidly oscillating phases $e^{i(\omega_{j}-\omega_{k})t}$ can be neglected, except of course when $\omega_{j}-\omega_{k}=0$.  This is valid at times longer than the reciprocal of the smallest energy gaps, $t \gg [ \min_{\omega_{j} \neq \omega_{k}}   |\omega_{j}-\omega_{k}|]^{-1} $.  For now, we assume this to be true, but later we will show that a much weaker energy gap condition entails many of the same features.  Making the SA leads to:
\begin{equation}
    X = - \sum_{j} \sum_{k; \omega_{k}= \omega_j}  \ket{j}\bra{k} \int_{0}^{\infty} e^{i \omega_{k}s}  f_{j,k}(s)   ds  
\end{equation}
We see that SA has removed any dependence on $t$ making time evolution Markovian.  Furthermore, the SA forces $X$ to commute with $H_{S}$, and so $X$ has the same block-diagonal structure as $H_S$.  

All matrices can be decomposed into $X=-H-P$ where $H^{\dagger}=-H$ and $P^{\dagger}=P$.  Performing just such a decomposition of $X$ we can show, via Bochner's theorem, that the Hermitian part $P$ has eigenvalues that are real and nonnegative (see App.~\ref{APPbochner}).  Furthermore, both matrices have real-valued elements, so $H^{\dagger}=H^T$ and $P^{\dagger}=P^T$, which entails that $H$ has purely imaginary eigenvalues, just as expected.  In Sec.~\ref{sec_Convergence} we saw that decoherence rates are governed by $P$. Recall that $P_G$ is the restriction of $P$ to the kernel of $H_S$ (equivalently the groundspace of the associated Hamiltonian, assuming $E_0=0$), and the decoherence rates between groundstates are governed by the spectrum of $P_G$.  Furthermore, this restricted $P_G$ matrix naturally emerges when one considers a relaxed SA assumption.

Recall that the validity of SA required that all energy gaps are large compared to a relevant time scale.  Many interesting topological systems have a degenerate groundspace with a large gap to the first excited state, but then the spectrum of excitations will be dense or even a continuum in the large system limit.  This means that SA cannot be used to eliminate transitions between different excited states.  However, provided the groundspace is gapped from excitations, we will have a limited application of SA that decouples $X$ into $O(X_{G} \oplus X_{E}) O^{T}$ and with $X_G$ describing the dynamics within the ground space.  Although SA will not apply to the dynamics $X_{E}$ of the excitations, we saw in Sec.~\ref{sec_Convergence} that decoherence in the groundspace is governed by only $X_G$.  In particular, it is dictated by the largest eigenvalue of $P_G$.

\section{Detailed balance and the Arrhenius law}
The bath temperature only influences what state we converge to, and not how quickly we get there. This conclusion is quite remarkable.  So much so, that naively it seems to violate some basic tenet of physics.  Two candidates are the Arrhenius law and detailed balance.  

The Arrhenius law is an empirical rule of thumb that has been successful in modeling chemical reactions.  Recently, it has been suggested that it may also apply to quantum memories, though violations have been observed in various settings~\cite{ku05,Shnirman05,yoshida14}.  The Arrhenius law predicts that decoherence times scale as $e^{\Delta /T}$ where $\Delta$ is the system gap.  We have a gapped degenerate ground space, but quasiparticles from the environment can poison the system without creating an excitation.  From this perspective perhaps we should consider $\Delta=0$, and our temperature independence to be consistent with the Arrhenius law.  However, although we have focused on ground space decoherence our observation apply also to the dynamics of excitations with a discrete spectrum.  That is, the rate at which excitation populations equilibrate is also temperature independent!  Indeed, the Arrhenius law is not a universal law and we conclude that it is absolutely violated in the domain of Gaussian fermions.

Detailed balance is a form of microscopic reversibility. It states that at thermal equilibrium the population transfer is symmetric for each process.  Consequently, the rate of transitions must depend on the temperature.  At first this seems to imply that decoherence rates must be temperature dependent. Indeed, detailed balance has been used to study decoherence times of spin (qubit) systems with the toric code Hamiltonian~\cite{alicki09} and cubic code Hamiltonian~\cite{bravyi13}.  These results revealed an exponential temperature dependence, and so one may expect this feature to be generic.  Though these are highly non-Gaussian systems.  In the Gaussian setting, Temme \textit{et. al.}~\cite{Temme14} have rigorously proved very general bounds on thermalization rates.  Their analysis appears to show an explicit temperature dependence, but closer inspection reveals variables that depend on the specific features of the bath.  These variables can be set, whilst still respecting detailed balance, to precisely cancel all temperature dependence.  We have seen that when the whole system-bath is Gaussian, this is indeed what happens. To further illustrate that temperature independence is consistent with detailed balance we present in App.~\ref{APPdb} a very simple classical Markov process where convergence rates decouple from temperature.  Detailed balance has an esteemed history going back to Boltzmann, who proclaimed it a key axiom of statistical mechanics and used it to great effect. However, there has been a recent surge of interest in non-equilibrium statistical mechanics, where detailed balance is violated, both in quantum~\cite{wichterich07,dorfman13} and classical settings~\cite{chou11}.  

\section{Comments on prior work}

Mazza \textit{et. al.}~\cite{Mazza13} studied the effect of various noise models on the 1-dimensional Kitaev chain.  Their main result is that various Hamiltonian perturbations will decohere the system, but this decoherence is suppressed by increasing the length of the Kitaev chain.  Mazza \textit{et. al.} conclude their paper by also discussing a more destructive noise model, open systems dynamics (see pg.4 of Ref.~\cite{Mazza13}).  This model is a master equation with the Hamiltonian of the standard Kitaev wire (with chemical potential set to zero), and Lindblad operators
\begin{equation}
    L_{\mu} = \eta a_{\mu}^{\dagger} = \eta \frac{1}{2} (c_{2\mu-1} + i c_{2\mu }) ,
\end{equation}
which allows for fermions hopping to the environment and where we use $\eta$ to parameterize the strength of the hopping.  They give numerical plots for decoherence of this model, but it can also be understood analytically.  We proceed by casting these master equations in the language of covariance matrices, and find\begin{equation}
    M = \frac{\eta^2}{4} \bigoplus_{j} \left(\begin{array}{cc} 1 & i \\
    -i & 1 \end{array} \right).
\end{equation}
Composing $M+M^*$ will cancel the imaginary parts, giving
\begin{equation}
    X =-H-(M+M^*)=-H - \frac{\eta^2}{2}  \id ,
\end{equation}
Note that the eigenvalues of $H$ are purely imaginary so that $e^{Ht}$ is unitary, but the dissipative component adds a constant real negative component.   We have that $D(t)=e^{Xt}= e^{-\eta^2 t / 2}e^{Rt}$ and $e^{X^{T}t}= e^{-\eta^2 t / 2}e^{H^{T}t}$.  Therefore, decoherence occurs at the rate $||D(t)||=e^{-\eta^2 t/2}$.  Here it is clear that system size and Hamiltonian gap play no role, provided $\eta$ is not a function of these variables.  Mazza \textit{et. al.} do not discuss how $\eta$ itself might depend on the energy gap or on temperature.  

Budich \textit{et. al.}~\cite{Budich12} made similar observations.  They considered two models where only the end of a Kitaev wire couples to the environment.  In both models the system-environment coupling has the form $H_{\mathrm{int}}= A_{\mathrm{bath}} c_{1}$ where $ A_{\mathrm{bath}}$ is some operator acting on the bath.  They also considered the standard Kitaev chain with zero chemical potential, so that the interaction commuted with the system Hamiltonian.  They observed rapid decoherence of Majorana edge modes, with no dependence on the energy gap.  These toy models are excellent ways to illustrate a serious deficit in prior claims to effectiveness of topological protection of Majorana edge modes.  However, they tell us little about what to expect when the interaction and Hamiltonian do not commute.    Furthermore, one may also wish to consider much more exotic models, such as Majorana fermions in 2D systems or even higher dimensions.  These gaps in previous work are now filled by the more general insights presented here.

\section{Conclusions and Acknowledgements}

We have seen that decoherence of Gaussian fermionic systems cannot be reduced by cooling.  For Markovian dynamics, we provided a microscopic derivation of the master equation in the weak coupling limit, leading to exponentially fast decoherence.  Therefore, to use Gaussian systems as a quantum memory they must be either highly non-Markovian or have minimal tunneling with any nearby Gaussian heat baths.  Eliminating tunneling is potentially challenging when in any of the many popular proposals for acquiring topological order through the proximity effect~\cite{Fu08,Sau10}.  In these proposals, electron hopping with an external $s$-wave superconductor is the mechanism by which topological robustness is acquired.  Both electron hopping and the superconducting Hamiltonian are Gaussian, and so this opens the door to temperature invariant decoherence.  For such systems it is urgent that we acquire a better understanding of the proximity effect from an open systems perspective. 

After completing this work, the author forwarded the manuscript to Leonardo Mazza who in return kindly shared several unpublished yet interesting results~\cite{Mazza12,Ippoliti14,Ippoliti14b}. These tackle related problems of interactions with fermionic baths, including various  numerical simulations of small fermionic (and bosonic) baths and numerous analytic insights. Particularly relevant is Sec. 7 of his PhD thesis~\cite{Mazza12}, where Mazza makes several observations also made here, although does he not remark on the temperature independence of decoherence rates.

We thank Pieter Kok and Keith Burnett for interesting discussions on Majorana fermions that lead to this research.  We thank Michael Kastoryano, Tomaz Prozen, Jens Eisert and Leonardo Mazza for comments on the manuscript.

\begin{appendix}

\section{Trace norm convergence}
\label{AppTRnorm}

We assume initial states of the form  $\rho(0)=\kb{\psi}{\psi}=\kb{\psi_G}{\psi_G} \otimes M_{G^{\perp}} $ and $\tilde{\rho}(0)=\kb{\phi}{\phi} \otimes M_{G^{\perp}} $ encoding different qubit states.  These initial states differ only on 4-Majorana modes within the groundspace, and the evolution of the covariance matrix shows that this property holds at later times so that 
\begin{equation}
\rho(t)=\rho_G(t) \otimes M_{G^{\perp}} \mbox{ , } \tilde{\rho}(t)=\tilde{\rho}_{G}(t) \otimes M_{G^{\perp}}.
\end{equation}
Therefore,
\begin{equation}
\begin{split}
    || \rho(t) - \tilde{\rho}(t) ||_{\mathrm{tr}} &=
     ||( \rho_G(t) - \rho_G'(t))\otimes M_{G^{\perp}} ||_{\mathrm{tr}} \\
     &= || \rho_G(t) - \rho_G'(t) ||_{\mathrm{tr}},
\end{split}
\end{equation}
where we have used $|| A \otimes B ||_{\mathrm{tr}}= || A   ||_{\mathrm{tr}} || B   ||_{\mathrm{tr}} $ and $|| M_{G^{\perp}} ||_{\mathrm{tr}} =1$.  The Hilbert space of 4 Majorana modes supports one qubit in the even parity subspace and one qubit in the odd parity subspace.  In other words, $ \rho_G(t) = \rho_G^{(0)}(t) \oplus \rho_G^{(1)}(t)$ and similarly $\tilde{\rho}_G(t) = \tilde{\rho}_G^{(0)}(t) \oplus \tilde{\rho}_G^{(1)}(t)$. Using $|| A \oplus B ||_{\mathrm{tr}}= || A   ||_{\mathrm{tr}}  +|| B   ||_{\mathrm{tr}}$ we have 
\begin{equation}
    || \rho(t) - \tilde{\rho}(t) ||_{\mathrm{tr}} = \sum_{x=0,1} || \rho_G^{(x)}(t)-\tilde{\rho}_G^{(x)}(t) ||_{\mathrm{tr}} .
\end{equation}
For a single qubit, we have $||\rho||_{\mathrm{tr}}=\mathrm{max}_{ \tilde{\rho} \in \mathcal{B}} \mathrm{tr}[  \tilde{\rho} \rho ]$ where the maximum is over all single qubit Hermitian unitary operators, such as the Pauli spin operators.  In a Majorana encoding the Pauli spin operators, indeed all single qubit Hermitian unitary operators, are quadratic observables.  These expectation values never exceed the operator norm of the $\delta(t)$. Therefore,
\begin{equation}
    || \rho(t) - \tilde{\rho}(t) ||_{\mathrm{tr}} \leq   || \delta(t) ||.\end{equation}
Finally, we make use of Eq.~(\ref{deltaCon}) to arrive at Eq.~(\ref{TRnormCon}).

\section{Remarks on detailed balance}
\label{APPdb}

Here we describe the concept of detailed balance for 2-state systems.  We show in this simple setting the concept is consistent with temperature invariant decoherence rates.  Furthermore, we show that Gaussian 2-mode Markov processes always obey this principle. Consider, a classical system with two possible states, with probabilities described by a Markov chain
\begin{equation}
    v = \left( \begin{array}{c} v_1 \\
   v_2 \end{array}   \right) = \left( \begin{array}{c} p \\
    1-p \end{array}   \right) .
\end{equation}
For simplicity we consider time to be in discrete steps, with a transition matrix
\begin{equation}
    P = \left( \begin{array}{cc} P_{1 \rightarrow 1} & P_{2 \rightarrow 1} \\
    P_{1 \rightarrow 2} & P_{2 \rightarrow 2}   \end{array} \right),
\end{equation}
so that $v(t)=P^tv$.  Conserving flow of probability requires $P_{k \rightarrow 1}+P_{k \rightarrow 2}=1$ for $k=1,2$.

We say $\pi$ is a stationary state of $P$, if $P \pi = \pi$, and where $\pi := (\alpha, 1- \alpha)$. The process $P$ satisfies detailed balance if
\begin{equation}
\label{EqDetBal}
    P_{1 \rightarrow 2} \alpha = P_{2 \rightarrow 1} (1-\alpha) .
\end{equation}
One can think of $\pi$ as a thermal distribution, so $\alpha=Z\exp(-E_1 \beta )$ and $(1-\alpha)=Z\exp(-E_2 \beta)$, where $Z = \exp(-E_1 \beta) + \exp(-E_2 \beta)$ is the partition function.  As usual, $\beta$ is inverse temperature.  In this thermal language, detailed balance entails that
\begin{equation}
   \frac{ P_{2 \rightarrow 1} }{P_{1 \rightarrow 2} } =\frac{\alpha}{1-\alpha}=\exp(\Delta \beta ) ,
\end{equation}
where $\Delta$ is the energy gap $E_2 - E_1$.  It appears that the (ratio of)  transition rates depend on the temperature of the steady state, and so one might be tempted to conclude that convergence rates likewise depend on temperature.  

The conservation of probability and detailed balance give 3 independent linear constraints on $P$, out of the 4 parameters of the matrix.  Therefore, the space of valid matrices is 1-dimensional and includes
\begin{equation}
    P_{\id} = \left( \begin{array}{cc} 1 & 0 \\
   0  & 1   \end{array} \right) ,
       P_{\pi} = \left( \begin{array}{cc}  \alpha & \alpha \\
   1-\alpha  & 1-\alpha   \end{array} \right).
\end{equation}
Both matrices satisfy Eq.~(\ref{EqDetBal}). Furthermore, for all Markov chains $v$ we have $P_{\id}v=v$ and $P_{\pi}v=\pi$.  The whole set of suitable matrices is contained in the span of these matrices,
\begin{equation}
\label{classicaleta}
\begin{split}
    P_{\eta} &= (1-\eta) P_{\id} + \eta P_{\pi}  \\
    &= \left( \begin{array}{cc}  \eta \alpha + (1-\eta) & \eta \alpha  \\
 \eta(1-\alpha)  &  \eta (1-\alpha)+ (1-\eta) \end{array} \right),
 \end{split}
\end{equation}
with $0 \leq \eta \leq \mathrm{min}[1/ \alpha ,  1 / (1- \alpha)]  \leq 2$ to ensure $P_{i \rightarrow j} \in [0,1]$.  Within these limits, $\eta$ is a free parameter.  We consider a general initial probability distribution it always has the form $v=\pi + p \tilde{\rho}$ for some value $p$, where $ \tilde{\rho} = (1,-1)$.  It is easy to confirm $P_{\id} \tilde{\rho} = \tilde{\rho}$ and $P_{\pi} \tilde{\rho} = 0$.  Therefore, $Pv=(1-\eta) v+ \eta\pi =\pi + p (1-\eta) \tilde{\rho}$ and for $t$ time steps this extends to
\begin{equation}
    v(t)=P^t v = \pi + p (1-\eta)^t \tilde{\rho} .
\end{equation}
This clearly shows exponentially rapid convergence to the equilibrium state at a speed governed by $\eta$.  More precisely, using any norm $||\ldots||$ to measure distance we have
\begin{equation}
    || v(t) - \pi || = p |1-\eta|^t ||\tilde{\rho} ||.
\end{equation}
The convergence speed is entirely independent of temperature, and only depends on the free parameter $\eta$.  The only temperature dependence lies in $\eta \leq \mathrm{min}[1/ \alpha ,  1 / (1- \alpha)]$, since $\alpha$ depends on temperature.  However, $\alpha \in [0,1]$ and so the range $\eta \in [0,1]$ is valid at all temperatures.  Therefore we can consider a family of Markov process with varying temperature and constant $\eta \in [0,1]$.  This family is consistent with detailed balance, but has a convergence rate independent of temperature.   

The convergence rate could vary with temperature if $\eta$ is a non-constant function of temperature. Although,  there is no reason \textit{a piori} to favour one function for $\eta$ over another.  Certainly, many possible temperature dependencies are consistent with detailed balance.  Unless, one has a physical model of the encompassing system and can perform a microscopic derivation of the Markov process, and so derive $\eta$.  This is exactly what we have performed for the case of Gaussian fermions, showing the analogous result of temperature independent $\eta$.  Lastly, we remark that this entire discussion can be recast in continuous time by considering $P$ to be the generator of a Markov process with transition matrix $Q(t)=e^{Pt}$.

It is still interesting to ask if Gaussian Markov processes obey detailed balance.  Let us just consider a pair of modes, with a 2-by-2 covariance matrix
\begin{equation}
    \Gamma = \left( \begin{array}{cc} 0 & -\lambda \\
 \lambda & 0
 \end{array}
 \right)
\end{equation}
The physical system is in one of two states (superpositions are disallowed by fermion parity superselection), with probability $p=(1+\lambda)/2$ and $1-p=(1-\lambda)/2$.  A Markov process maps $\Gamma \rightarrow X \Gamma X^T + Y$ where $Y$ is skew-symmetric.  Under this process, we find some $x,y$ such that $\lambda \rightarrow x \lambda + y$. Therefore, $p \rightarrow px + \frac{1}{2}(1+y-x)$ and  the probability transition matrix has the form
\begin{equation}
\label{Transition}
    P = \left( \begin{array}{cc} \frac{1}{2}(1+x+y) & \frac{1}{2}(1-x+y) \\
 \frac{1}{2}(1-x-y) & \frac{1}{2}(1+x-y)
 \end{array}
 \right)
\end{equation}
In the steady state $p_{\mathrm{ss}}= p_{\mathrm{ss}}x + \frac{1}{2}(1+y-x)$ and so that
\begin{align*}
    p_{\mathrm{ss}}&=\frac{1-x+y}{2(1-x)}, \\
    1-p_{\mathrm{ss}}&=\frac{1-x-y}{2(1-x)}.
\end{align*}
Therefore,
\begin{equation}
\frac{p_{\mathrm{ss}}}{1-p_{\mathrm{ss}}}=\frac{1-x+y}{1-x-y} = \frac{P_{2 \rightarrow 1}}{P_{1 \rightarrow 2}}
\end{equation}
so that detailed balance is satisfied for all $x,y$.

\section{Application of Bochner's theorem }
\label{APPbochner}

\end{appendix}

Here we show that the Hermitian part of $X$ has strictly negative eigenvalues.  The proof makes use of Bochner's theorem, which relates properties of functions to their Fourier transform.   To introduce this theorem, we first define the concept of functions of positive-type
\begin{defin}
An absolutely integrable function $\phi : \mathbb{C} \rightarrow \mathbb{C}$ is of positive type if for all sets of complex numbers $\{ c_{1}, c_{2},\ldots  \}$ the following summation is real-valued and positive
\begin{equation}
    \sum_{n,m} c^*_{n} c_{m}  \phi( c_{n} - c_{m} ) \geq 0 .
\end{equation}
\end{defin}
Note that being of positive type is very different from a function taking positive values. Now we can state
\begin{theorem}
\textit{Bochner's theorem: Let $\phi$ be an absolutely integrable function.  The fourier transformed function $\tilde{\phi}$ is a real-valued positive function if and only if $\phi$ is of positive type.}
\end{theorem}
Returning to the problem at hand, $B=(X+X^{\dagger})/2$ is Hermitian by construction.  From Eq.~(\ref{Eq_X_RWAed}) we already have an expression for $X$, and considering $X^{\dagger}$ we observe that
\begin{eqnarray} \nonumber
    X^{\dagger} &=& - \sum_{j} \sum_{k, \omega_{k}=\omega_j} \ket{k}\bra{j} \int_{0}^{\infty} e^{i \omega_{k} s} f_{j,k}^{*}(s) ds , \\
   &=&  \sum_{j} \sum_{k, \omega_{k}=\omega_j} \ket{k}\bra{j} \int_{0}^{-\infty} e^{i \omega_{k} s} f_{j,k}^{*}(-s) ds , 
\end{eqnarray}
where we have made the change of variables $s \rightarrow -s$.  Switching the order of integration,
\begin{equation}
    X^{\dagger} =- \sum_{j} \sum_{k, \omega_{k}=\omega_j}  \ket{k}\bra{j} \int_{-\infty}^{0} e^{i \omega_{k} s} f_{j,k}^{*}(-s) ds . \\
\end{equation}
Next, we use that $f_{j,k}^{*}(-s)= f_{k,j}(s)$, which can be seen from
\begin{eqnarray} \nonumber
    f_{j,k}^*(-s) &=& \bra{j}H_{I}^{\dagger}e^{H_{B}s}H_{I} \ket{k}^{\dagger}, \\
    & = & \bra{k}H_{I}^{\dagger}e^{-H_{B}^{\dagger}s}H_{I} \ket{j}  ,
\end{eqnarray}
and using $H_{B}^{\dagger}=H_{B}^T=-H_{B}$ we have the result.  Applying this to our expression for $X^{\dagger}$, and switching the dummy variables $j \leftrightarrow k$ gives 
\begin{equation}
    X^{\dagger} =-  \sum_{j} \sum_{k, \omega_{k}=\omega_j} \ket{j}\bra{k}  \int_{-\infty}^{0} e^{i \omega_{k} s} f_{j,k}(s) ds .\\ 
\end{equation} 
This differs from $X$ in only the domain of the integral and so
\begin{equation} \nonumber
    X^{\dagger}+X = -\sum_{j} \sum_{k, \omega_{k}=\omega_j} \ket{j}\bra{k}  \int_{-\infty}^{\infty} e^{i \omega_{k} s} f_{j,k}(s) ds .   
\end{equation} 
For each set of variables, $j,k$, the integral is a Fourier transform of $f_{j,k}$ evaluated at $\omega_{k}$, so
\begin{equation} \nonumber
    X^{\dagger}+X = -\sum_{j} \sum_{k, \omega_{k}=\omega_j} \ket{j}\bra{k}  \tilde{f}_{j,k}(s) ds .   
\end{equation} 
Notice, we have denoted Fourier transforms with a tilde. Next, we show that $f_{j,k}$ is of positive-type. For all $\{ c_{1}, c_{2}, \ldots \}$
\begin{equation}
   \sum_{n,m}  c_{n}^*  c_{m} f_{j,k}( c_{n} - c_{m})= \bk{w}{w}    ,
\end{equation} 
where 
\begin{equation}
   \ket{w} = \sum_{m} c_{m} e^{H_{B}c_{m}} H_I \ket{k}.
\end{equation}
Since $\bk{w}{w} \geq 0$, we can apply Bochner's theorem and conclude that all $\tilde{f}_{j,k}(\omega_{k})$ are positive and real.  If $H_{S}$ is a nondegenerate matrix, then there would be no multiplicity of eigenvectors with the same eigenvalue and $-\tilde{f}_{k,k}(\omega_{k})$ would represent the real-negative eigenvalues of $X+X^{\dagger}$.  However, for degenerate matrices there is a freedom of choice in the basis $\{ \ket{k} \}$, but we can always set this to be the eigenbasis of $X+X^{\dagger}$.  Therefore, $X+X^{\dagger}$ has real negative eigenvalues.

\end{document}